\documentstyle[twoside,psfig]{article}
\begin{document}
\centerline{\bf MARKET DEPTH AND PRICE DYNAMICS: A NOTE}

\bigskip
\centerline{ FRANK H. WESTERHOFF}

\bigskip
\centerline{\it University of Osnabrueck,
Department of Economics}

\centerline{\it Rolandstrasse 8, D-49069 Osnabrueck, Germany}

\centerline{\it e-mail: fwesterho@oec.uni-osnabrueck.de}
\baselineskip=10pt
\vspace*{0.225truein}

\bigskip
\vspace*{0.25truein} Abstract: This note explores the consequences of
nonlinear price impact functions on price dynamics within the
chartist-fundamentalist framework. Price impact functions may be nonlinear
with respect to trading volume. As indicated by recent empirical studies,
a given transaction may cause a large (small) price change if market depth
is low (high). Simulations reveal that such a relationship may create
endogenous complex price fluctuations even if the trading behavior of
chartists and fundamentalists is linear.

\vspace*{5pt} Keywords: Econophysics; Market Depth; Price Dynamics;
Nonlinearities; Technical and Fundamental Analysis.

\section{Introduction} 
\vspace*{-0.5pt} \noindent Interactions between heterogeneous agents,
so-called chartists and fundamentalists, may generate endogenous price
dynamics either due to nonlinear trading rules or due to a switching
between simple linear trading rules.$^{1,2}$ Overall, multi-agent models
appear to be quite successful in replicating financial market
dynamics.$^{3,4}$ In addition, this research direction has important
applications. On the one hand, understanding the working of financial
markets may help to design better investment strategies.$^{5}$ On the
other hand, it may facilitate the regulation of disorderly markets. For
instance, Ehrenstein shows that the imposition of a low transaction tax
may stabilize asset price fluctuations.$^{6}$

Within these models, the orders of the traders typically drive the price
via a log linear price impact function: Buying orders shift the price
proportionally up and selling orders shift the price proportionally down.
Recent empirical evidence suggests, however, that the relationship between
orders and price adjustment may be nonlinear. Moreover, as reported by
Farmer et al., large price fluctuations occur when market depth is
low.$^{3,7}$ Following this observation, our goal is to illustrate a novel
mechanism for endogenous price dynamics.

We investigate -- within an otherwise linear chartist-fundamentalist setup
-- a price impact function which depends nonlinearly on market depth. To
be precise, a given transaction yields a larger price change when market
depth is low than when it is high. Simulations indicate that such a
relationship may lead to complex price movements. The dynamics may be
sketched as follows. The market switches back and forth between two
regimes. When liquidity is high, the market is relatively stable. But low
price fluctuations indicate only weak trading signals and thus the
transactions of speculators decline. As liquidity decreases, the price
responsiveness of a trade increases. The market becomes unstable and price
fluctuations increase again.

The remainder of this note is organized as follows. Section 2 sketches the
empirical evidence on price impact functions. In section 3, we present our
model, and in section 4, we discuss the main results. The final section
concludes.

\section{Empirical Evidence}
\noindent
Financial prices are obviously driven by the orders of heterogeneous
agents. However, it is not clear what the true functional form of price
impact is. For instance, Farmer proposes a log linear price impact
function for theoretical analysis while Zhang develops a model with
nonlinear price impact.$^{8,9}$ His approach is backed up by empirical
research that documents a concave price impact function. According to
Hasbrouck, the larger is the order size, the smaller is the price impact
per trade unit.$^{10}$ Also Kempf and Korn, using data on DAX futures, and
Plerou et al., using data on the 116 most frequently traded US stocks,
find that the price impact function displays a concave curvature with
increasing order size, and flattening out at larger values.$^{11,12}$
Weber and Rosenow fitted a concave function in the form of a power law and
obtained an impressive correlation coefficient of 0.977.$^{13}$ For a
further theoretical and empirical debate on the possible shape of the
price impact function with respect to the order size see Gabaix et al.,
Farmer and Lillo and Plerou et al.$^{14-16}$

But these results are currently challenged by an empirical study which is
crucial for this note. Farmer et al. present evidence that price
fluctuations caused by individual market orders are essentially
independent of the volume of the orders.$^{7}$ Instead, large price
fluctuations are driven by fluctuations in liquidity, i.e. variations in
the market's ability to absorb new orders. The reason is that even for the
most liquid stocks there can be substantial gaps in the order book. When
such a gap exists next to the best price -- due to low liquidity -- even a
small new order can remove the best quote and trigger a large price
change. These results are supported by Chordia, Roll and Subrahmanyam who
also document that there is considerable time variation in market wide
liquidity and Lillo, Farmer and Mantenga who detect that higher
capitalization stocks tend to have smaller price responses for the same
normalized transaction size.$^{17,18}$

Note that the relation between liquidity and price impact is of direct
importance to investors developing trading strategies and to regulators
attempting to stabilize financial markets. Farmer et al. argue, for
instance, that agents who are trying to transact large amounts should
split their orders and execute them a little at a time, watching the order
book, and taking whatever liquidity is available as it enters.$^{7}$
Hence, when there is a lot of volume in the market, they should submit
large orders. Assuming a concave price impact function would obviously
lead to quite different investment decisions. Ehrenstein, Westerhoff and
Stauffer demonstrate, for instance, that the success of a Tobin tax
depends on its impact on market depth.$^{19}$ Depending on the degree of
the nonlinearity of the price impact function, a transaction tax may
stabilize or destabilize the markets.

\section{The Model}
\noindent Following Simon, agents are boundedly rational and display a
rule-governed behavior.$^{20}$ Moreover, survey studies reveal that
financial market participants rely strongly on technical and fundamental
analysis to predict prices.$^{21,22}$ Char\-tists typically extrapolate
past price movements into the future. Let $P$ be the log of the price.
Then, their orders may be expressed as

\begin{equation}
D^C_t = a(P_t -P_{t-1}),
\end{equation}
where $a$ is a positive reaction coefficient denoting the strength of the
trading. Accordingly, technical traders submit buying orders if prices go
up and vice versa. In contrast, fundamentalists expect the price to track
its fundamental value. Orders from this type of agent may be written as

\begin{equation}
D^F_t = b(F-P_t).
\end{equation}
Again, $b$ is a positive reaction coefficient, and $F$ stands for the log
of the fundamental value. For instance, if the asset is overvalued,
fundamentalists submit selling orders.

As usual, excess buying drives the price up and excess selling drives it
down so that the price adjustment process may be formalized as

\begin{equation}
P_{t+1} = P_t + A_t(wD^C_t + (1-w)D^F_t),
\end{equation}
where $w$ indicates the fraction of chartists and $(1-w)$ the fraction of
fundamentalists. The novel idea is to base the degree of price adjustmen
$A$ on a nonlinear function of the market depth.$^{23}$ Exploiting that 
given excess demand has a larger (smaller) impact on the price if the
trading volume is low (high), one may write

\begin{equation}
A_t = \frac{c}{(|wD^C_t|+|(1-w) D^F_t|)^d}.
\end{equation}
The curvature of $A$ is captured by $d\geq 0$, while $c>0$  is a shift
parameter.

For $d=0$, the price adjustment function is log-linear.$^{1,3}$ In that
case, the law of motion of the price, derived from combining (1) to (4),
is a second-order linear difference equation which has a unique steady
state at

\begin{equation}
P_{t+1} = P_t = P_{t-1} = F.
\end{equation}
Rewriting Schur's stability conditions, the fixed point is stable for

\begin{equation}
0<c<\left\{\begin{array}{ll}
\displaystyle\frac{1}{aw} & \mbox{for~~} w> \displaystyle \frac{b}{4a +b}\\
\displaystyle \frac{2}{b(1-w)-2aw}\qquad & \mbox{else}
\end{array}.\right.
\end{equation}
However, we are interested in the case where $d>0$. Combining (1)-(4) and
solving for $P$ yields

\begin{equation}
P_{t+1} = P_t + c \frac{wa(P_t - P_{t-1}) + (1-w)
b(F-P_t)}{(|wa(P_t-P_{t-1})|+|(1-w)b(F-P_t)|)^d},
\end{equation}
which is a two-dimensional nonlinear difference equation. Since (7)
precludes closed analysis, we simulate the dynamics to demostrate that the
underlying structure gives rise to endogenous deterministic motion.

\section{Some Results}
\noindent
Figure 1 contains three bifurcation diagrams for $0<d<1$ and $w=0.7$
(top), $w=0.5$ (central) and $w=0.3$ (bottom). The other parameters are
fixed at $a=b=c=1$  and the log of the fundamental value is $F=0$. We
increase $d$ in 500 steps. In each step, $P$ is plotted from
$t=$1001-1100. Note that bifurcation diagrams are frequently used to
illustrate the dynamic properties of nonlinear systems.

\begin{figure}[htbp] 
\vspace*{13pt}
\centerline{\psfig{file=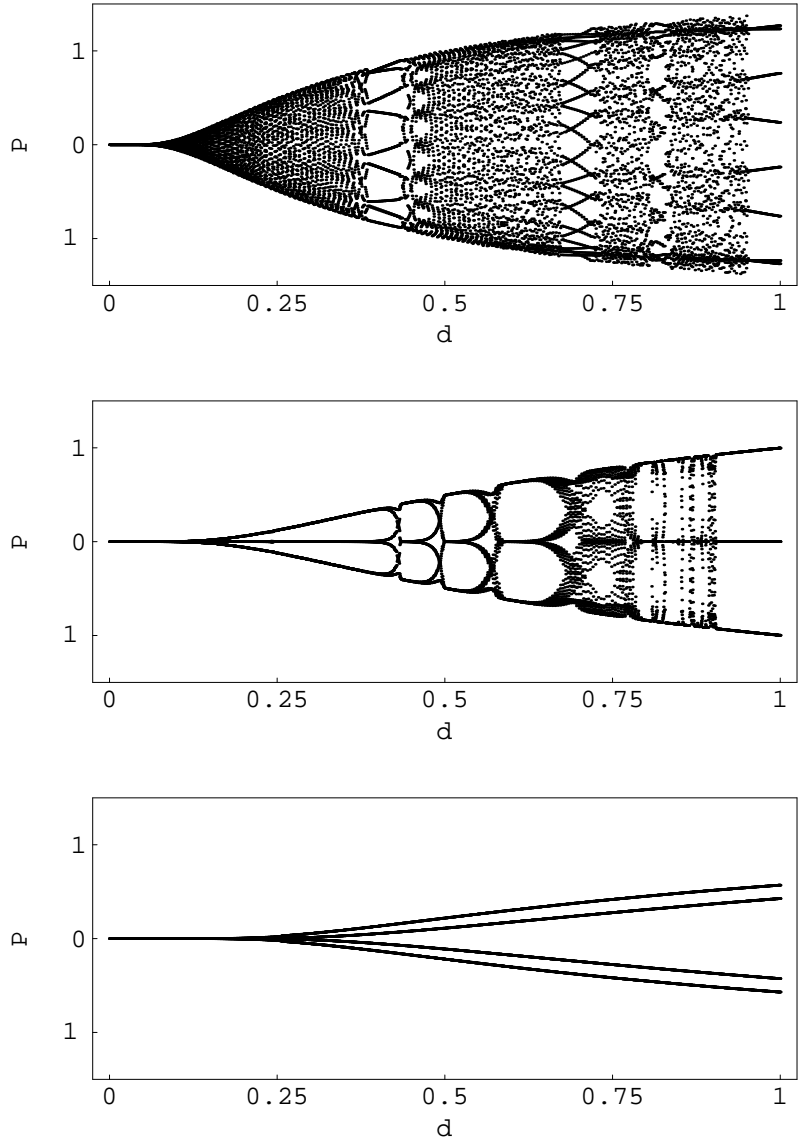}} 
\vspace*{13pt}
\caption{Bifurcation diagrams for $0<d<1$ and $w=0.7$ (top), $w=0.5$
(central) and $w=0.3$ (bottom). The other parameters are fixed at
$a=b=c=1$. The parameter $d$ is increased in 500 steps. For each value of
$d$, $P$ is plotted from $t=$1001-1100. The log of the fundamental value
is $F=0$.}
\end{figure}

Figure 1 suggests that if $d$ is small, there may exist a stable
equilibrium. For instance, for $w=0.5$, prices converge towards the
fundamental value as long as $d$ is smaller than around 0.1. If $d$ is
increased further, the fixed point becomes unstable. In addition, the
range in which the fluctuations take place increases too. Note also that
many different types of bifurcation occur. Our model generates the full
range of possible dynamic outcomes: fixed points, limit cycles, quasi
periodic motion and chaotic fluctuations. For some parameter combinations
coexisting attractors emerge. Comparing the three panels indicates that
the higher the fraction of chartists, the less stable the market seems to
be.

To check the robustness of endogenous motion, figure 2 presents
bifurcation diagrams for $0 < a <2$ (top), $0< b < 2$ (central) and $0< c
< 2$ (bottom), with the remaining parameters fixed at $a=b=c=1$ and
$d=w=0.5$. Again, complicated movements arise. While chartism seems to
destabilize the market, fundamentalism is apparently stabilizing.
Naturally, a higher price adjustment destabilizes the market as well.
Overall, many parameter combinations exist which trigger complicated
motion.\footnote{To observe permanent fluctuations only small variations
in $A$ are needed. Suppose that $A$ takes two values centered around the
upper bound of the stability condition $X$, say $X-Y$ and $X+Y$, depending
on whether trading volume is above or below a certain level $Z$. Such a
system obviously produces nonconvergent but also nonexplosive fluctuations
for arbitrary values of $Y$ and $Z$.}

\begin{figure}[htbp] 
\vspace*{13pt}
\centerline{\psfig{file=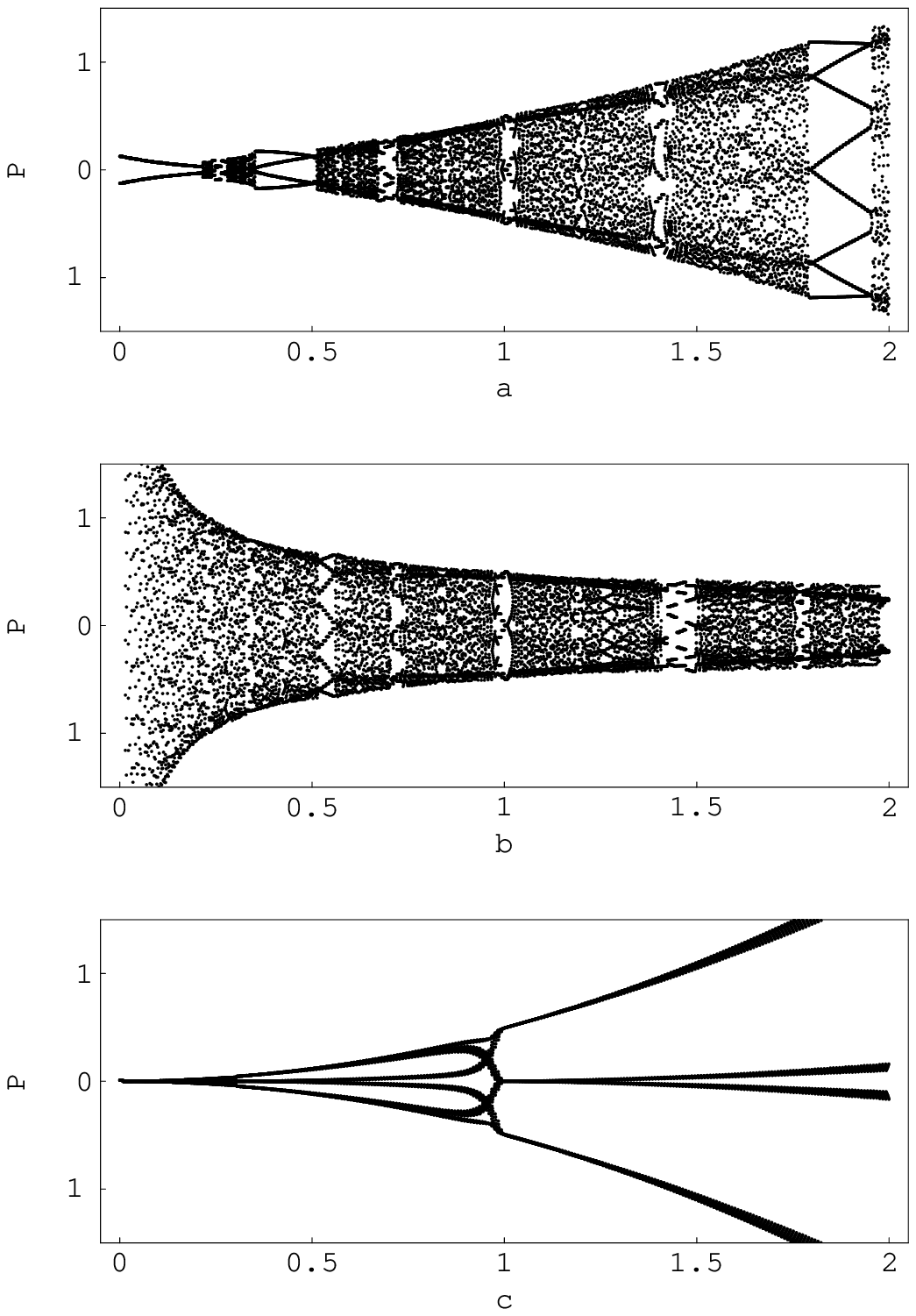}} t
\vspace*{13pt}
\caption{Bifurcation diagrams for $0<a<2$ (top), $0<b<2$ (central) and
$0<c<2$ (bottom), with the remaining parameters fixed at $a=b=c=1$ and
$d=w=0.5$. The bifurcation parameters are increased in 500 steps. For each
value, $P$ is plotted from $t=$1001-1100. The log of the fundamental value
is $F=0$.}
\end{figure}

Let us finally explore what drives the dynamics. Figure 3 shows the
dynamics in the time domain for $a=0.85$, $b=c=1$, and $d=w=0.5$. The
first, second and third panel present the log of the price $P$, the price
adjustment $A$ and the trading volume $V$ for 150 observations,
respectively. Visual inspection reveals that the price circles around its
fundamental value without any tendency to converge. Nonlinear price
adjustment may thus be an endogenous engine for volatility and trading
volume. Note that when trading volume drops the price adjustment increases
and price movements are amplified. However, the dynamics does not explode
since a higher trading volume leads again to a decrease in the price
adjustment.

\begin{figure}[htbp] 
\vspace*{13pt}
\centerline{\psfig{file=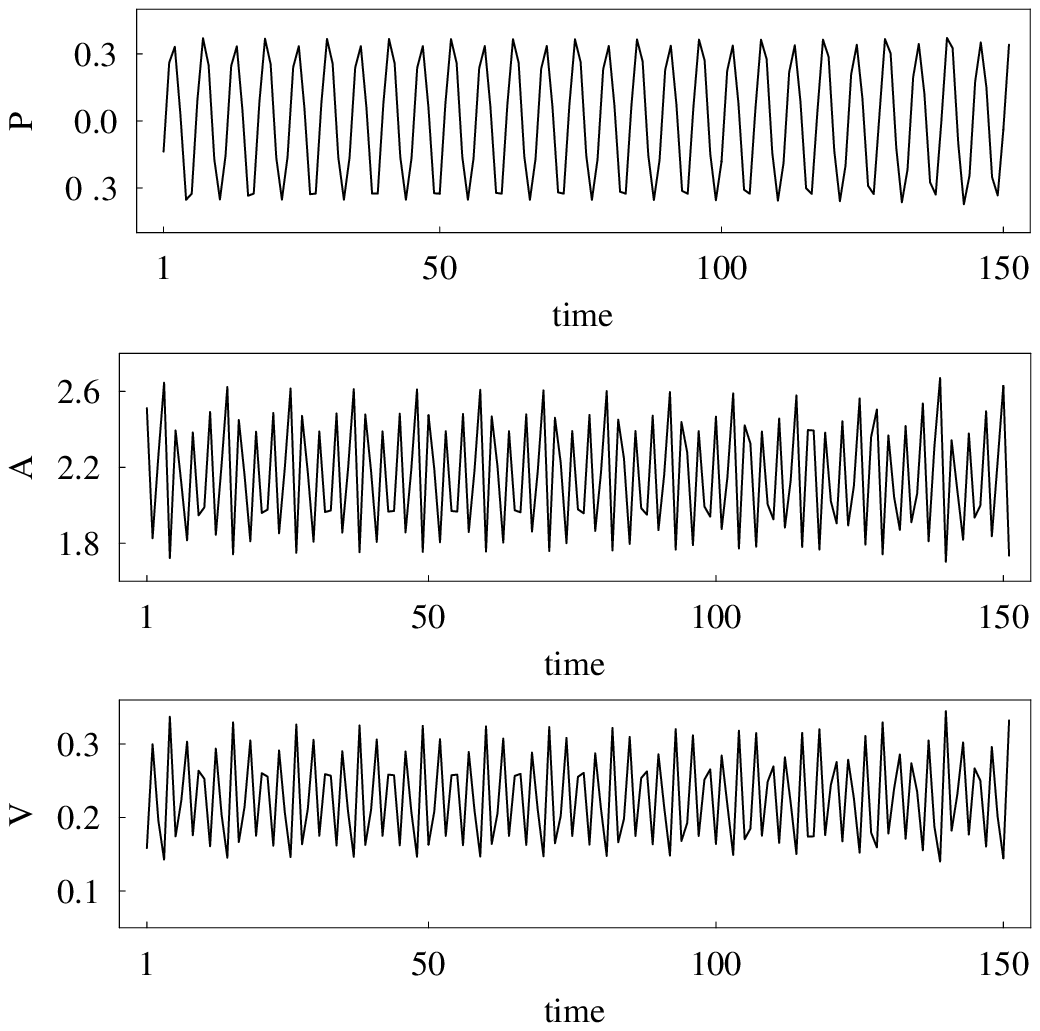}} 
\vspace*{13pt}
\caption{The dynamics in the time domain for $a=0.85,~b=c=1$, and
$d=w=0.5$. The first, second and third panel show the price $P$, the price
adjustment $A$ and the trading volume $V$ for 150 observations,
respectively. The log of the fundamental value is $F=0$.}
\end{figure}

Finally, figure 4 displays the price (top panel) and the trading volume
(bottom panel) for 5000 observations $(a = 0.25,~b = 1,~c = 50,~d = 2$ and
$w=0.5)$. As can be seen, the dynamics may become quite complex. Remember
that trading volume increases with increasing price changes (orders of
chartists) and/or increasing deviations from fundamentals (orders of
fundametalists). In a stylized way, the dynamics may thus be sketched as
follows: Suppose that trading volume is relatively low. Since the price
adjustment $A$ is strong, the system is unstable. As the trading becomes
increasingly hectic, prices start to diverge from the fundamental value.
At some point, however, the trading activity has become so strong that, 
due to the reduction of the price adjustment $A$, the system becomes
stable. Afterwards, a period of convergence begins until the system jumps
back to the unstable regime. This process continually repeats itself but
in an intricate way.

\begin{figure}[htbp] 
\vspace*{13pt}
\centerline{\psfig{file=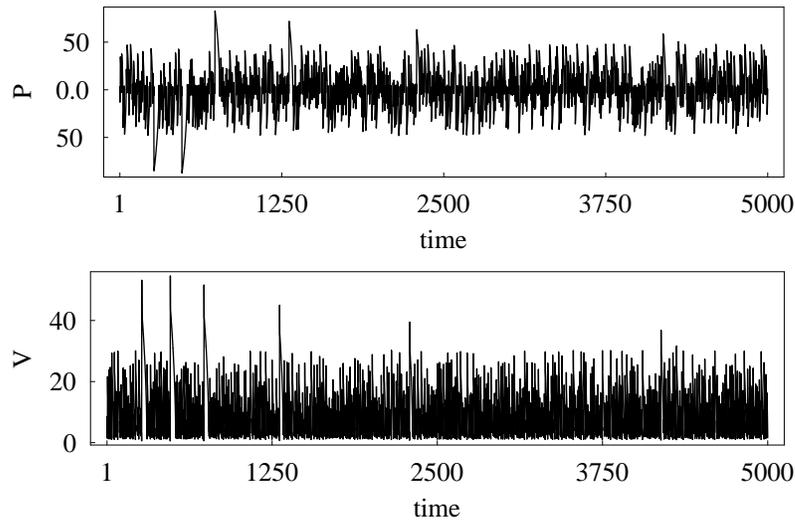}} 
\vspace*{13pt}
\caption{The dynamics in the time domain for $a=0.25,~b=1,~c=50,~d=2$ and
$w=0.5$. The first (second) panel displays the price $P$ (the trading
volume $V$) for 5000 observations. The log of the fundamental value is
$F=0$.}
\end{figure}

\section{Conclusions}
\noindent
When switching between simple linear trading rules and/or relying on
nonlinear strategies, interactions between heterogeneous agents may cause
irregular dynamics. This note shows that changes in market depth also
stimulate price changes. The reason is that if market liquidity goes down,
a given order obtains a larger price impact. For a broad range of
parameter combinations, erratic yet deterministic trajectories emerge
since the system switches back and forth between stable and unstable
regimes.

\newpage

{\bf References}

\begin{enumerate}
\itemsep-1ex
\item D. Farmer and S. Joshi, {\it Journal of Economic Behavior and
Organizations} {\bf 49}, 149
(2002).
\item T. Lux and M. Marchesi, {\it International Journal of Theoretical
and Applied Finance} {\bf 3}, 675 (2000).
\item R. Cont and J.-P. Bouchaud, {\it Macroeconomic Dynamics} {\bf 4}, 170 (2000).
\item D. Stauffer, {\it Advances in Complex Systems} {\bf 4}, 19 (2001).
\item D. Sornette and W. Zhou, {\it  Quantitative Finance} {\bf 2}, 468 (2002).
\item G. Ehrenstein, {\it International Journal of Modern Physics C} {\bf 13}, 1323 (2002).
\item D. Farmer, L. Gillemot, F. Lillo, S. Mike and A. Sen, What Really
Causes Large Price Changes?, SFI Working Paper, 04-02-006, 2004.
\item D. Farmer, {\it Industrial and Corporate Change} {\bf 11}, 895 (2002).
\item Y.-C. Zhang, {\it Physica A} {\bf 269}, 30 (1999).
\item J. Hasbrouck, {\it Journal of Finance} {\bf 46}, 179 (1991).
\item A. Kempf and O. Korn, {\it  Journal of Financial Markets} {\bf 2}, 29 (1999).
\item V. Plerou, P. Gopikrishnan, X. Gabaix and E. Stanley, {\em Physical
Review E} {\bf 66},
027104, 1 (2002).
\item P. Weber and B. Rosenow,  Order Book Approach to Price Impact,
Preprint cond-mat/0311457, 2003.
\item X. Gabaix, P. Gopikrishnan, V. Plerou and E. Stanley, {\it Nature} {\bf 423,} 267 (2003).
\item D. Farmer and F. Lillo, {\it  Quantitative Finance} {\bf 4}, C7 (2004).
\item V. Plerou, P. Gopikrishnan, X. Gabaix and E. Stanley, {\it Quantitative Finance} {\bf 4},
C11 (2004).
\item T. Chordia, R. Roll and A. Subrahmanyam, {\it Jour\-nal of Finance} {\bf 56}, 501 (2001).
\item F. Lillo, D. Farmer and R. Mantegna, {\it Nature} {\bf 421}, 129 (2003).
\item G. Ehrenstein, F. Westerhoff and D. Stauffer D., Tobin Tax and
Market Depth, Preprint cond-mat/0311581, 2001.
\item H. Simon, {\it Quarterly Journal of Economics} {\bf 9}, 99 (1955).
\item M. Taylor and  H. Allen, {\it Journal of International Money and Finance} {\bf 11}, 304 (1992).
\item Y.-H. Lui and D. Mole, {\it  Journal of International Money and
Finance} {\bf 17}, 535 (1998).
\item D. Sornette and K. Ide, {\it International Journal of Modern Physiscs C} {\bf 14}, 267  (2003).
\end{enumerate}

\end{document}